\documentstyle[multicol,aps,prl,ifthen,epsfig]{revtex}
\tighten

\newcommand{\wide}[2]{                                                        %
\end{multicols}                                                               %
\widetext                                                                     %
\noindent                                                                     %
\ifthenelse{\equal{#1}{t}}                                                    %
{}                                                                            %
{                                                                             %
\raisebox{0.1in}[0in][0.02in]{$\rule{3.575in}{0.002in}                        %
\rule{0.002in}{0.08in}$}                                                      %
}                                                                             %
#2                                                                            %
\ifthenelse{\equal{#1}{b}}                                                    %
{}                                                                            %
{                                                                             %
{\raisebox{-0.1in}[0in][0.02in]                                               %
{\hspace{3.575in}$\rule{0.002in}{0.08in}                                      %
\rule[0.08in]{3.575in}{0.002in}$}                                             %
}                                                                             %
}                                                                             %
\begin{multicols}{2}                                                          %
\noindent                                                                     %
}                                                                             %

\begin{document}


\title{Non-existence of spin-asymmetry of conductance in two-terminal devices}

\author{P. Bruno and T.P. Pareek}

\address{Max-Planck-Institut f\"ur Mikrostrukturphysik,
Weinberg 2, D-06120 Halle, Germany}


\maketitle


\begin{multicols}{2}


In a recent Letter \cite{Grundler2001}, Grundler considered a
two-terminal device consisting of a junction between a
ferromagnetic metal (FM) and a two-dimensional electron gas
(2DEG) with Rashba effect (see Fig.~1). Grundler claims that the
conductance of the junction would change when the magnetization
of the FM (taken parallel to the $y$ axis) is reversed. We show in
the present Comment that this conclusion is wrong and is based
upon a misconception of the nature of the Rashba effect. On the
other hand, we also show that the conductance changes when the
magnetization is rotated from the $y$ axis to the $z$ axis.

The main argument against Grundler's claim is based upon the
B\"uttiker symmetry relations for electrical conduction
\cite{Buttiker1986}. B\"uttiker's theory establishes that a the
resistance (or conductance) of a phase-coherent two-terminal
device is invariant upon reversal of an external magnetic field.
This is a consequence of the unitarity of the scattering matrix,
which itself results from the principle of charge conservation.
B\"uttiker also showed that this conclusion remains valid when
the device is connected to phase-breaking reservoirs (such that
the current flowing from or to the reservoirs vanishes); this
implies that B\"uttiker's symmetry relation remains valid for an
{\em incoherent} two-terminal device. One can easily show that, if
some part of the device is magnetically ordered, B\"uttiker's
result remains valid, provided both the magnetic field and the
magnetization are reversed. This results holds even in the
presence of spin-orbit interaction. In order to obtain a change
of the conductance upon magnetization reversal, the device must
have at least three terminals \cite{Bruno1997,Pareek2001}.

Grundler's result therefore violates B\"uttiker's symmetry
relation. This is due to Grundler's incorrect assumption that the
Rashba effect acts ``like an effective magnetic field $B^\star$
parallel to the $y$ axis''. The Hamiltonian of a 2DEG with Rashba
interaction is $H=\hbar^2 ({k_x}^2 + {k_y}^2)/2m^\star + \alpha
(k_y \sigma_x - k_x \sigma_y )$ \cite{Bychkov1984}. As can be seen
from Fig.~1, the spin-splitting due to the Rashba effect differs
considerably from the one due to the Zeeman effect in a field
$B^\star$. Further, Grundler wrongly claims that the
spin-splitted subbands have different populations and Fermi
velocities, which is obviously incorrect.

We have also performed some numerical calculations of the
conductance of FM/2DEG junctions, by using a Green's function
formulation of the Landauer-B\"uttiker formula \cite{Pareek2001}.
The results are shown in Fig.~1 (bottom panel). It clearly
appears that the conductance is invariant upon magnetization
reversal, both in the ballistic and diffusive cases. On the other
hand, a significant conductance anisotropy (dependence on the
magnetization axis, but not on its sign) is clearly seen. This
effect does not violate B\"uttiker's symmetry relation and can be
obtained with a two-terminal device as shown in
Ref.~\cite{Pareek2001}. It is of second order with respect to the
spin-orbit and exchange interactions and can be viewed as a
variant of the anisotropic magnetoresistance of ferromagnets,
which is due to the simultaneous presence of exchange splitting
and spin-orbit interaction; the interesting aspect here is that
the effect occurs although the exchange and spin-orbit
interactions are spatially separated.

\begin{figure}
\centering \epsfig{file=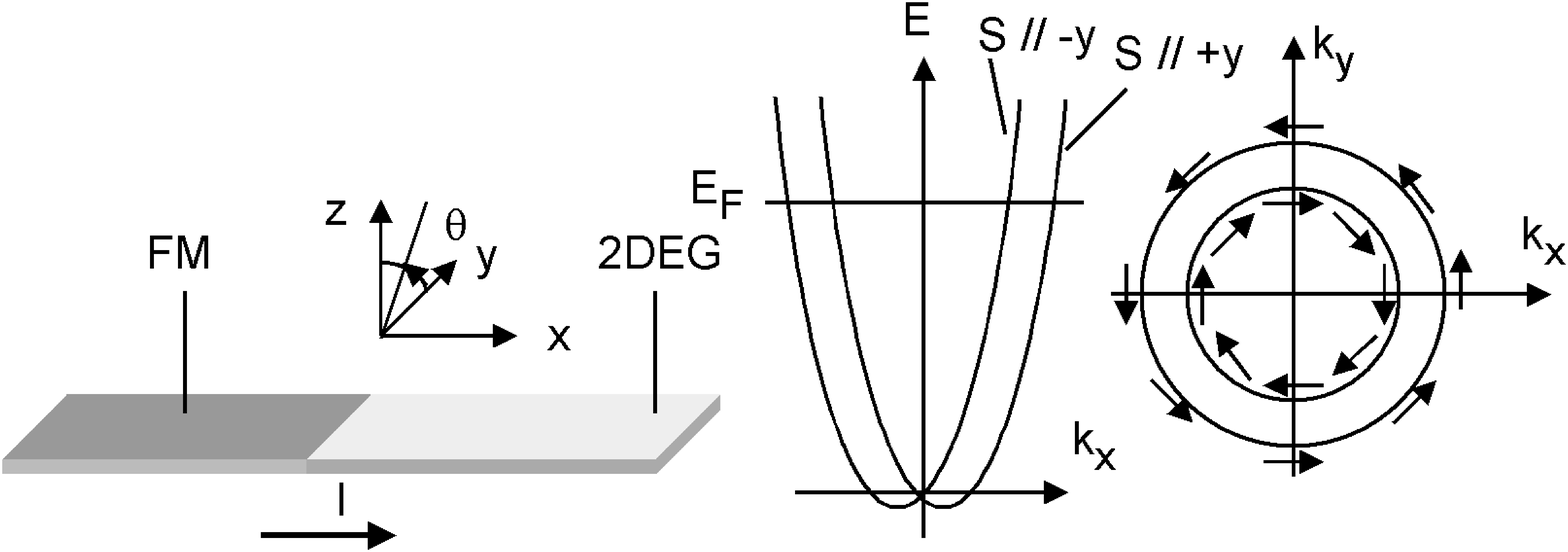, width=8cm} \vspace{0.25cm}
\vspace{0.0cm} \centering \epsfig{file=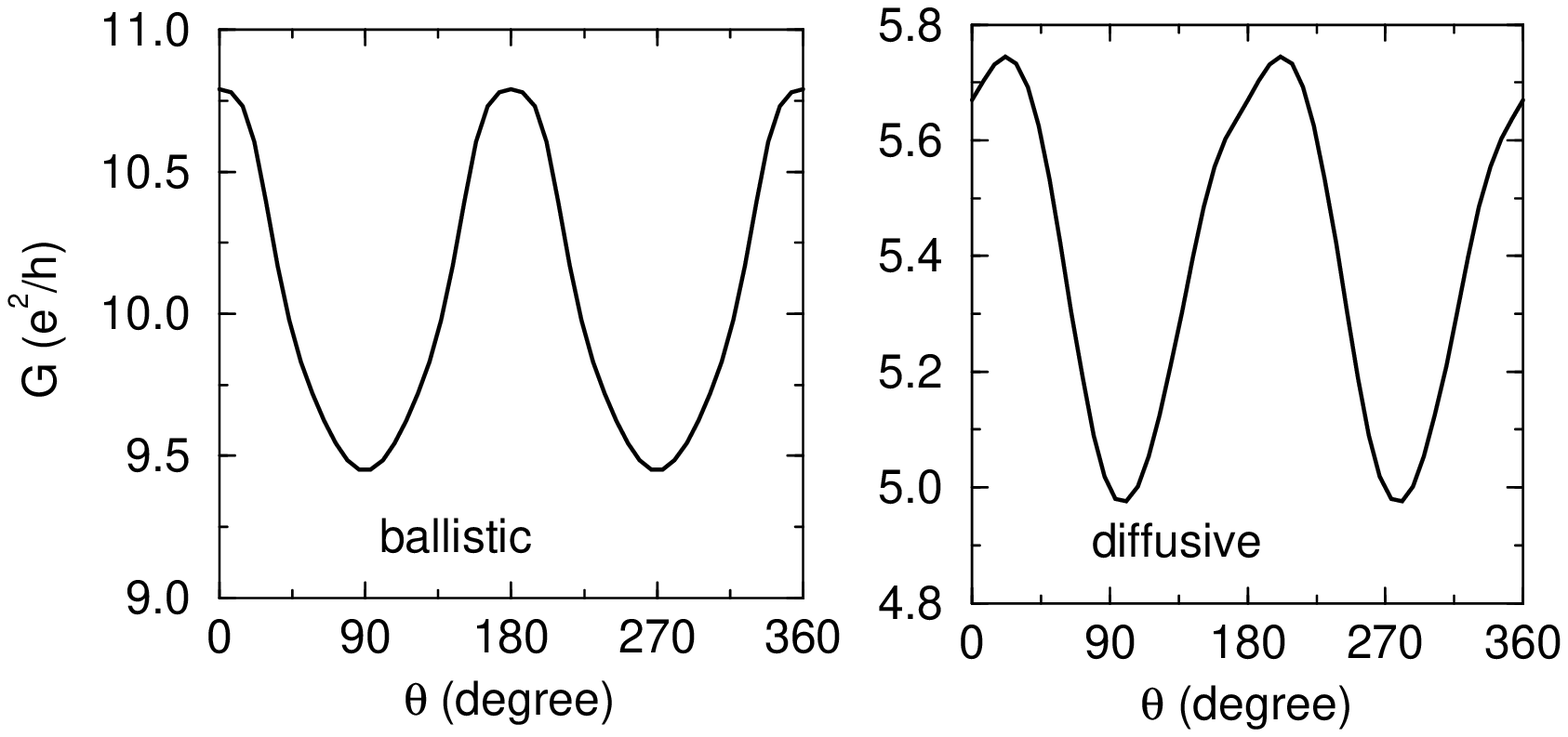, width=8cm}
\caption{Top: schematic description of the device geometry
(left), and of the band-structure (center) and Fermi surface
(right) of a 2DEG with Rashba effect; the arrows indicate the
direction of the spin-quantization axis. Bottom: Two-terminal
conductance of a FM/2DEG junction as a function of the angle
$\theta$ of the magnetization in $yz$ plane for the ballistic
(left) and diffusive (right) cases. Calculations were performed
on a 50$\times$50 lattice with $k_Fa=1$, FM exchange splitting
$\Delta$ given by $\Delta /\varepsilon_F =0.5$, 2DEG Rashba
parameter $k_F\alpha /\varepsilon_F = 0.03$; for the diffusive
case (no configuration averaging), the mean-free-path $l$ is
given by $k_Fl=30$.}
\end{figure}
%

\end{multicols}

\end{document}